\documentclass[10pt,twocolumn,letterpaper]{article}

\usepackage{iccv}
\usepackage{times}
\usepackage{epsfig}
\usepackage{graphicx}
\usepackage{amsmath}
\usepackage{amssymb}
\usepackage{verbatim}
\usepackage{caption}
\usepackage{booktabs}
\usepackage{multirow}
\usepackage{pifont}
\usepackage[accsupp]{axessibility}
\newcommand{\cmark}{\ding{51}}
\newcommand{\xmark}{\ding{55}}
\usepackage[symbol]{footmisc}

\usepackage[breaklinks=true,bookmarks=false]{hyperref}

\iccvfinalcopy 

\def\iccvPaperID{10421} 

\ificcvfinal\fi

\let\oldtwocolumn\twocolumn
\renewcommand\twocolumn[1][]{%
    \oldtwocolumn[{#1}{
    \begin{center}
           \includegraphics[scale=0.42]{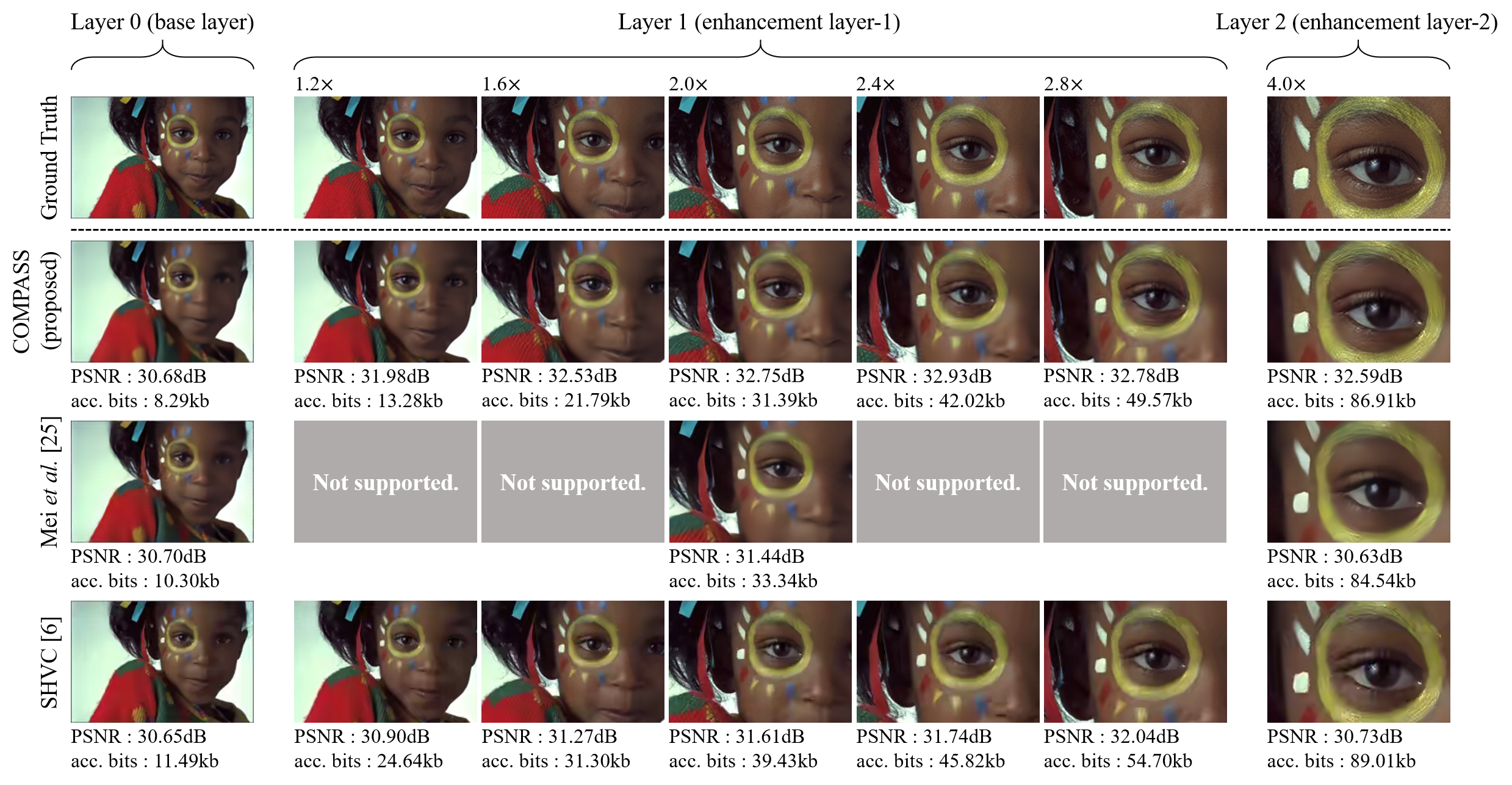}
           \vspace{-0.8cm}
           \captionof{figure}{\textbf{Visual comparison results of spatially scalable image compression methods for arbitrary scale factors.} The image of Layer 2 is reconstructed from the image of Layer 1 with scale factor 2.0$\times$ for each method. The `acc. bits' indicates the accumulated bits up to the corresponding layers.}
           \label{fig:figure_page1}
        \end{center}
    }]
}

\begin{document}

\title{COMPASS: High-Efficiency Deep Image Compression with Arbitrary-scale Spatial Scalability}

\author{Jongmin Park\\
KAIST\\
{\tt\small jm.park@kaist.ac.kr}
\and
Jooyoung Lee\\
ETRI\\
{\tt\small leejy1003@etri.re.kr}
\and
Munchurl Kim\thanks{Corresponding author.}\\
KAIST\\
{\tt\small mkimee@kaist.ac.kr}
}

\maketitle
\ificcvfinal\thispagestyle{empty}\fi

\begin{abstract}
\vspace{-0.505cm}
Recently, neural network (NN)-based image compression studies have actively been made and has shown impressive performance in comparison to traditional methods. However, most of the works have focused on non-scalable image compression (single-layer coding) while spatially scalable image compression has drawn less attention although it has many applications. In this paper, we propose a novel NN-based spatially scalable image compression method, called COMPASS, which supports arbitrary-scale spatial scalability. Our proposed COMPASS has a very flexible structure where the number of layers and their respective scale factors can be arbitrarily determined during inference. To reduce the spatial redundancy between adjacent layers for arbitrary scale factors, our COMPASS adopts an inter-layer arbitrary scale prediction method, called LIFF, based on implicit neural representation. We propose a combined RD loss function to effectively train multiple layers. Experimental results show that our COMPASS achieves BD-rate gain of -58.33\% and -47.17\% at maximum compared to SHVC and the state-of-the-art NN-based spatially scalable image compression method, respectively, for various combinations of scale factors. Our COMPASS also shows comparable or even better coding efficiency than the single-layer coding for various scale factors.
\end{abstract}

\section{Introduction}
\label{sec:intro}
Recently, image compression has become increasingly important with the growth of multimedia applications.
The exceptional performance of neural network (NN)-based methods in computer vision has led to active research on NN-based image compression methods \cite{toderici2017full, balle2016end, theis2017lossy, balle2018variational, minnen2018joint, lee2018context, cheng2020learned, lee2019end, liu2019non, minnen2020channel, he2021checkerboard, agustsson2019generative, mentzer2020high}, resulting in remarkable improvements in coding efficiency.
However, although the same content is often consumed in various versions in multimedia systems, most existing NN-based image compression methods must separately compress an image into multiple bitstreams for their respective versions, thus leading to low coding efficiency.
To resolve this issue, there have been a few recent studies~\cite{Toderici16, su2020scalable, jia2019layered, zhang2019learned, guo2019deep, ma2022deepfgs, Lu_progressive_2021, mei2021learning} on NN-based scalable image compression, where various versions of an image are encoded into a single bitstream in a hierarchical manner with multiple layers. Each layer is in charge of en/decoding one corresponding version of the image, and typically, redundancy between adjacent layers is reduced by a prediction method for higher coding efficiency. 

The scalable coding methods are divided into two classes: quality scalable codecs for the images of different quality levels and spatially scalable codecs for the images of different sizes. In this paper, we focus on the spatially scalable coding that has not been actively studied compared with the quality scalable coding. Upon our best knowledge, only one previous study~\cite{mei2021learning} deals with the spatially scalable coding in the recent deep NN-based approach.

In conventional tool-based scalable coding, SVC~\cite{schwarz2007overview} and SHVC~\cite{boyce2015overview} have been standardized by MPEG~\cite{le1991mpeg} for video coding standards, as extensions to H.264/AVC~\cite{wiegand2003overview} and H.265/HEVC~\cite{sullivan2012overview}, respectively. Despite of significant coding efficiency improvement compared with separate single-layer compression of different versions (simulcast coding), the scalable coding has not yet been widely adopted for real-world applications \cite{boyce2015overview, seregin2014common}. One reason may be lower coding efficiency of the accumulated bitstream for the larger version compared with the single-layer coding of the same size.
The scalable coding often yields lower coding efficiency due to its insufficient redundancy removal capability between the layers.

In addition, for the existing NN-based method~\cite{mei2021learning}, only one fixed scale factor 2 is used between adjacent layers as shown in Figure \ref{fig:figure_page1}. This limitation makes it not practical for real-world applications that require a variety of scale combinations.
For example, an image of 4,000$\times$2,000 size needs to be encoded into SD (720$\times$480), HD (1,280$\times$720) and FHD (1,920$\times$1,080) versions which are not in powers of 2 scales compared to the input size.
Therefore, in order to support for the one-source-multiple-use (OSMU) with spatially scalable image compression, it is worthwhile for spatially scalable image compression to support arbitrary scale factors between the different layers.

To address the aforementioned issues, we propose a novel NN-based image \underline{COMP}ression network with \underline{A}rbitrary-scale \underline{S}patial \underline{S}calability, called COMPASS. Our COMPASS supports spatially scalable image compression that encodes multiple arbitrarily scaled versions of an image into a single bitstream in which each version of the image is encoded with its corresponding layer. Inspired by LIIF~\cite{chen2021learning} and Meta-SR~\cite{hu2019meta}, we adopt an inter-layer arbitrary scale prediction method in the COMPASS, called Local Implicit Filter Function (LIFF), based on implicit neural representation that can effectively reduce the redundancy between adjacent layers and also supports arbitrary scale factors. In addition, it should be noted that our COMPASS exploits only one shared prediction/compression module for all the enhancement layers, thus it effectively provides the extensibility in terms of the number of layers and also reduces the number of model parameters. For effective and stable optimization of the hierarchically recursive architecture of COMPASS, we introduce a combined RD loss function.

Based on its superior inter-layer prediction capability, our COMPASS significantly improves the coding efficiency compared to the existing scalable coding methods~\cite{boyce2015overview, mei2021learning}, and achieves comparable or even better coding efficiency compared to the single-layer coding for various scale factors. Note that the coding efficiency of the single-layer coding has been regarded as the upper bound of the scalable coding efficiency.
Furthermore, to the best of our knowledge, our COMPASS is the first NN-based spatially scalable image compression method that supports arbitrary scale factors with high coding efficiency. Our contributions are summarized as:

\begin{itemize}\itemsep1pt
\item The COMPASS is the first NN-based spatially scalable image compression method for \textit{arbitrary} scale factors.
\item The COMPASS adopts an inter-layer arbitrary scale prediction, called LIFF, which is based on implicit neural representation to reduce redundancy effectively as well as to support the arbitrary scale factors. Additionally, we propose a combined RD loss function to effectively train multiple layers.
 \item Our COMPASS significantly outperforms the existing spatially scalable coding methods \cite{boyce2015overview, mei2021learning}. Furthermore, to the best of our knowledge, the COMPASS is the first work that shows comparable or even better performance in terms of coding efficiency than the single-layer coding for various scale factors, based on a same image compression backbone.
\end{itemize}

\section{Related Work}
\label{sec:related}
\noindent \textbf{Neural Network-based Image Compression.}
Recently, there have been proposals to optimize neural network (NN)-based image compression methods in an end-to-end manner.
Toderici~\etal \cite{toderici2017full} first proposed a deep convolutional NN-based image compression method, while Ball\'e~\etal \cite{balle2016end} and Theis~\etal \cite{theis2017lossy} adopted the entropy model-based approaches that jointly minimize the rate and distortion terms in the optimization phase.
Subsequent models, such as hyperprior~\cite{balle2018variational}, auto-regressive models~\cite{minnen2018joint, lee2018context}, Gaussian Mixture Models \cite{cheng2020learned, lee2019end}, non-local attention modules \cite{liu2019non}, channel-wise auto-regressive entropy models \cite{minnen2020channel} and the checkerboard context model~\cite{he2021checkerboard}, have improved coding efficiency.
There are also a few generative model-based studies~\cite{agustsson2019generative, mentzer2020high} for human perception-oriented compression. Recently, several NN-based variable-rate compression models~\cite{choi2019variable, Cui2020, rippel2021elf, song2021variable, Lu2021, lee2022selective} have been studied to support the multiple compression quality levels with a single trained model. Despite the significant improvements in coding efficiency and functionality brought about by the NN-based image compression networks, there remains an issue with coding efficiency when encoding different versions of an image as described in Sec. \ref{sec:intro}.
\newline

\noindent \textbf{Spatially Scalable Image Compression.}
For OSMU applications that supports various-sized display devices, images often need to be compressed and transmitted to target devices with appropriate spatial sizes.
To meet this requirement, the scalable extensions of traditional coding standards, H.264/AVC~\cite{wiegand2003overview} and H.265/HEVC~\cite{sullivan2012overview} have been developed as SVC~\cite{schwarz2007overview} and SHVC~\cite{boyce2015overview}, respectively.
Recently, NN-based approaches for scalable image compression \cite{Toderici16, su2020scalable, jia2019layered, zhang2019learned, guo2019deep, ma2022deepfgs, Lu_progressive_2021, mei2021learning} have also been proposed.
However, most of these works focus on quality scalability and only Mei \etal \cite{mei2021learning} deals with spatial scalability.
Mei \etal \cite{mei2021learning} proposed a hierarchical architecture which outperforms the simulcast coding and SVC \cite{schwarz2007overview}, and shows comparable performance with SHVC \cite{boyce2015overview} in terms of coding efficiency.
However, it can only support fixed integer scale factors with powers of 2.
Moreover, they didn't provide any experimental evidence on the extended multiple enhancement layers more than 2, although they proposed the layer extension concept.
\newline

\noindent \textbf{Arbitrary Scale Super-Resolution.}
With the advancement of neural networks, several recent works have proposed super-resolution with arbitrary scale factors, such as \cite{hu2019meta, chen2021learning, fu2021residual, wang2021learning, xu2021ultrasr}.
Hu \etal \cite{hu2019meta} introduced Meta-SR, the first neural network-based method for super-resolution with arbitrary scales.
In Meta-SR, the Meta-Upscale module takes the relative coordinate and scale factor as input to dynamically predict the upscaling filters.
Wang \etal \cite{wang2021learning} proposed an asymmetric super-resolution method using conditional convolution.
Cheng \etal \cite{chen2021learning} presented a continuous image representation method with Local Implicit Image Function (LIIF), and achieved outstanding performance for large scale ($\times{30}$) super-resolution which is out of training distribution.
Xu \etal \cite{xu2021ultrasr} used periodic encoding with the implicit function.
Inspired by these arbitrary scale super-resolution methods \cite{chen2021learning, hu2019meta}, we adopt them for the inter-layer arbitrary scale prediction in our COMPASS. We refer to this method as the Local Implicit Filter Function (LIFF), which can effectively reduce redundancy between adjacent layers with arbitrary scale factors.

\section{Proposed Method}
\label{sec:proposed}

\begin{figure}
\centering
\includegraphics[scale=0.34]{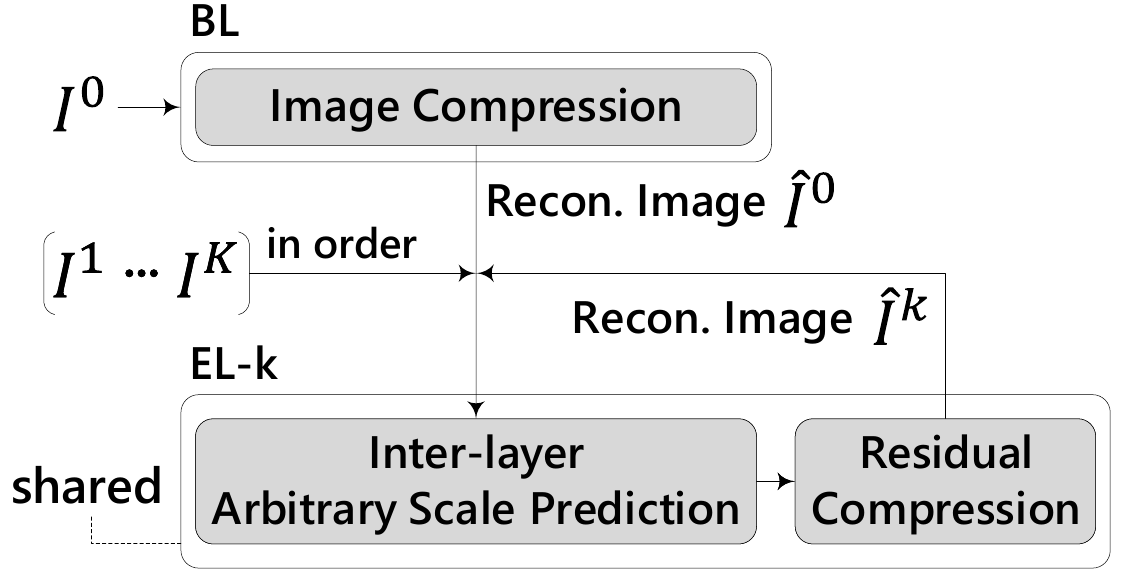}
\caption{The COMPASS supports spatially scalable coding of $K$+1 arbitrary scaled versions of an image using a base layer (BL) and one or more enhancement layers (ELs).
The EL-$k$ ($1 \leq k \leq K$) exploits a shared subnetwork that the Inter-layer Arbitraty Scale Prediction and Residual Compression modules.
$I_0$ indicates the smallest-sized input image in the BL. $I_1, ..., I_K$ are the input images in the ELs in an increasing order of scale factors where $I_K$ is the largest-sized input image. Note that the scale factor between two adjacent layers can be any arbitrarily positive value.}
\vspace{-0.4cm}
\label{fig:concept}
\end{figure}

\begin{figure*}
\centering
\includegraphics[scale=0.6]{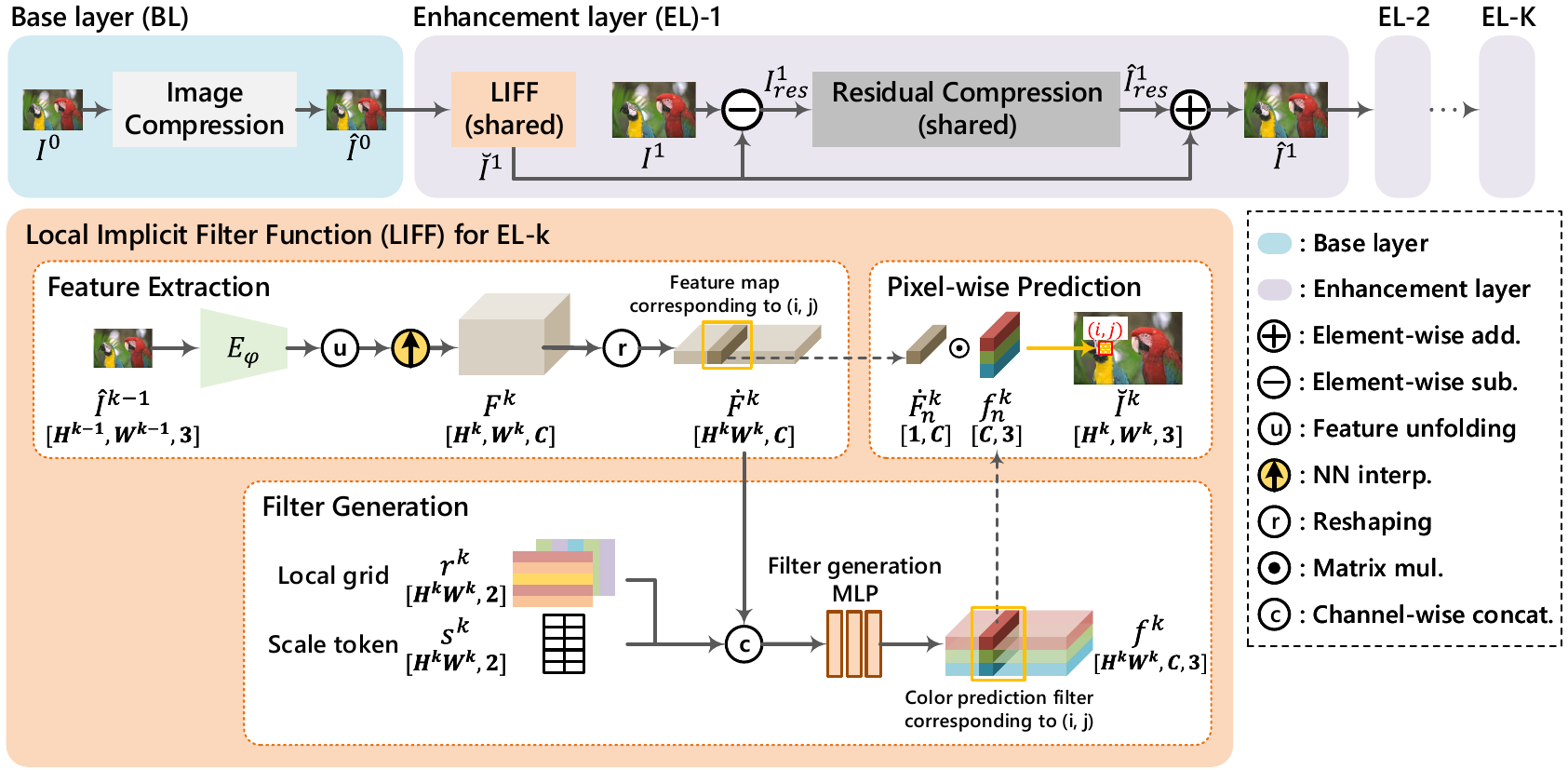}
\caption{\textbf{Overall architecture of our COMPASS.} It consists of a base layer (BL) depicted in the sky blue box and one or more enhancement layers (ELs) depicted in the light purple boxes which operate in an iterative manner. Note that we exploit the shared modules (LIFF and residual compression) for multiple ELs.}
\vspace{-0.4cm}
\label{fig:overall_architecture}
\end{figure*}


\subsection{Overall architecture}
\label{subsec:overall_architecture}
Figure \ref{fig:concept} depicts a flow diagram of our COMPASS.
The COMPASS comprises of two types of layers: a base layer (BL) that encodes the lowest resolution image, and one or more enhancement layers (ELs) that sequentially encode multiple higher resolution images of arbitrary scales.
For spatially scalable coding of ($K$+1)-scaled images $\{I^{0}, ..., I^{K}$\} of gradually increasing sizes with arbitrary scale factors, the COMPASS operates with multiple coding in the BL and $K$ ELs, each of which encodes the correspondingly scaled input image. It should be noted that the COMPASS exploits the shared modules for all the ELs, each of which recursively operates as depicted in Figure \ref{fig:concept}.
In the BL, the smallest-sized input image $I^0$ is fed into a CNN-based image compression module to reconstruct $\hat{I}^0$. In the EL-$k$, the corresponding input image $I^k$ and the reconstructed image $\hat{I}^{k-1}$ of the previous layer are fed into the current enhancement layer to reconstruct $\hat{I}^k$. Specifically, in the EL-$k$, the inter-layer arbitrary scale prediction module can effectively estimate and reduce the spatial redundancy between $\hat{I}^{k-1}$ and $I^k$ for arbitrary scale factor. Therefore, the residual compression module only encodes the resulting essential residues in reconstructing $\hat{I}^k$ with high coding efficiency. Figure \ref{fig:overall_architecture} shows the overall architecture of our COMPASS. 
We describe the operation of COMPASS with $K$+1 layers as:
\vspace{-0.1cm}
\begin{equation}
\hat{I}^k =
\begin{cases}
IC(I^k), & \mbox{if }k=0 \;\; (\text{BL})\\
\breve{I}^k + \hat{I}_{res}^k, & \mbox{if }k>0 \;\; (\text{EL-}k)\\  
\end{cases}
\vspace{-0.1cm}
\end{equation}
where, for $k>0$,
\vspace{-0.1cm}
\begin{equation}
\begin{aligned}
\breve{I}^k = \psi(\hat{I}^{k-1}, \boldsymbol{s}^k, \boldsymbol{r}^k) \;\; \text{and} \;\; \hat{I}_{res}^k = RC(I_{res}^k),
\end{aligned}
\vspace{-0.1cm}
\end{equation}
\noindent where $IC(\cdot)$ refers to an image compression module of the BL and $RC(\cdot)$ refers to a residual compression module of the EL-$k$, as shown in Figure \ref{fig:overall_architecture}.
We adopt the Mean-scale~\cite{minnen2018joint} architecture for both the compression modules.
$\breve{I}^k \in \mathbb{R}^{H^k \times W^k \times 3}$ refers to an arbitrarily upscaled prediction for the EL-$k$, and it is predicted from the smaller reconstruction $\hat{I}^{k-1}$ by the LIFF module which is denoted as $\psi(\cdot)$, and $I_{res}^k$ indicates a residual image between $I^k$ and $\breve{I}^k$.
The LIFF module takes a local grid $\boldsymbol{r}^k \in \mathbb{R}^{H^kW^k \times 2}$ and a scale token $\boldsymbol{s}^k \in \mathbb{R}^{H^kW^k \times 2}$ as additional inputs, which are described in details in Sec. \ref{subsec:LIFF}.
Since the output of convolutional layers are progressively reduced in half due to the convolution of stride 2 in the encoder part, the input to the encoder part is often padded into the size of a power of 2 in a lump at the beginning. This actually deteriorates the coding efficiency in our image compression with arbitrary scale factors. Therefore, we adopt a convolutional-layer-wise padding scheme where (i) a replicate padding with the padding size of 1 is performed if the width or height size of the input is an odd number in each convolutional layer of the encoder part of the residual compression module; and (ii) we crop out the padded region for the output of the corresponding convolutional layer of the decoder part.

\subsection{LIFF: Inter-layer arbitrary scale prediction}
\label{subsec:LIFF}
To achieve high coding efficiency with the COMPASS, it is essential to effectively reduce the redundancy between adjacent layers.
For this, we adopt an inter-layer arbitrary scale prediction method using a local implicit filter function (LIFF) which is based on the local implicit image function (LIIF) \cite{chen2021learning} and Meta-SR \cite{hu2019meta}. Our LIFF module first transforms the reconstruction $\hat{I}^{k-1}$ of the previous layer into the feature domain and then increases its resolution to match the arbitrarily upscaled prediction $\breve{I}^k$ through a simple interpolation.
Our LIFF module also generates the color prediction filter for each pixel coordinate and then estimate the RGB color pixel-wise by applying the generated filter to the extracted feature slice corresponding to the target pixel coordinate.
The procedure of the LIFF module is divided into 3 stages: 1) Feature Extraction, 2) Filter Generation, 3) Pixel-wise Prediction, as illustrated in the orange box of Figure \ref{fig:overall_architecture}.

\noindent \textbf{Feature Extraction.}
We extract feature information from the reconstruction $\hat{I}^{k-1}$ of the previous layer through an RDN-like feature extractor $E_\varphi$~\cite{zhang2018residual}, and apply feature unfolding \cite{chen2021learning} and nearest-neighbor upsampling to generate the feature map $\boldsymbol{F}^k \in \mathbb{R}^{H^{k} \times W^{k} \times C}$.

\noindent \textbf{Filter Generation.}
We generate the color prediction filter $\boldsymbol{f}^k \in \mathbb{R}^{H^{k}W^{k} \times C \times 3}$ using a filter generation MLP as
\vspace{-0.3cm}
\begin{equation}
\label{eq:CAFWP}
    \begin{aligned}
    \boldsymbol{f}^k = \phi(\lceil \dot{\boldsymbol{F}}^{k}, \boldsymbol{r}^k, \boldsymbol{s}^k \rfloor; \theta),
    \end{aligned}
\vspace{-0.3cm}
\end{equation}
where $\dot{\boldsymbol{F}}^{k} \in \mathbb{R}^{H^kW^k \times C}$ refers to the flattened feature map,
$\phi(\cdot)$ refers to the filter generation MLP with parameters $\theta$, and $\lceil \cdot \rfloor$ refers to channel-wise concatenation. The local grid $ \boldsymbol{r}^k \in \mathbb{R}^{H^{k}W^{k} \times 2}$ and the scale token $ \boldsymbol{s}^k \in \mathbb{R}^{H^{k}W^{k} \times 2}$ follow the same process as in the LIIF \cite{chen2021learning}. The local grid $\boldsymbol{r}^k$ is a normalized relative coordinate between the reconstruction $\hat{I}^{k-1}$ of the previous layer and the upscaled prediction $\breve{I}^k$, which is formulated as $\boldsymbol{r}^k(i,j) = \boldsymbol{p}^k(i,j) - \boldsymbol{p}^{k-1}(i',j')$.
$\boldsymbol{p}^k(i,j)$ refers to a normalized coordinate of the upscaled prediction $\breve{I}^k$ at pixel coordinate $(i,j)$, and $\boldsymbol{p}^{k-1}(i',j')$ indicates a corresponding normalized coordinate of the reconstruction $\hat{I}^{k-1}$ of the previous layer at pixel coordinate $(i',j')$.
We adopt the nearest-neighbor to find the pixel correspondence.
The normalized coordinate is calculated as $\boldsymbol{p}^l(i,j) = [-1 + (2i+1)/{H^l}, -1 + (2j+1)/{W^l}]$, where $i \in [0, H^l-1]$ and $j \in [0, W^l-1].$
The scale token $ \boldsymbol{s}^k$ indicates the height/width ratio between $\hat{I}^{k-1}$ and $\breve{I}^k$. $ \boldsymbol{s}^k$ then contains all the same ratio values of $(2\cdot H^{k-1}/H^k, 2\cdot W^{k-1}/W^k)$.

\begin{figure}[!]
\centering
\includegraphics[scale=0.37]{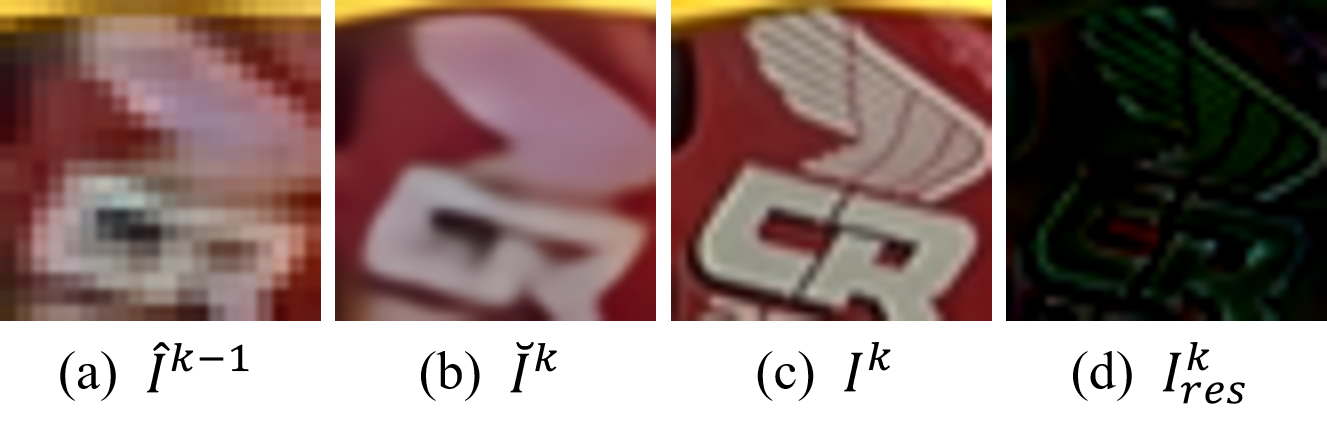}
\vspace{-0.7cm}
\caption{A predicted image via the LIFF module. (a) the reconstruction of the previous layer $k$-1, (b) the output (predicted image) of the LIFF module, (c) the input image of the current layer $k$, (d) the residual image as the input of the residual compression module.}
\vspace{-0.4cm}
\label{fig:LIFF_input}
\end{figure}

\noindent \textbf{Pixel-wise Prediction.}
To determine the RGB color of the arbitrarily upscaled prediction $\breve{I}^k$ at pixel coordinate $(i,j)$, we apply the color prediction filter $\boldsymbol{f}^{k}_n$ for the generated feature map $\dot{\boldsymbol{F}}^k_n$ by a simple matrix multiplication as
\vspace{-0.3cm}
\begin{equation}
\label{eq:mapping}
    \begin{aligned}
    \breve{I}^k(i,j) = \dot{\boldsymbol{F}}^{k}_{n} \odot \boldsymbol{f}^{k}_n,
    \end{aligned}
\vspace{-0.3cm}
\end{equation}
where $n \in [0, H^kW^k-1]$ indicates the batch index number which is corresponding to the pixel coordinate $(i, j)$ of the prediction $\breve{I}^k$ via $n = i + j\cdot H^k$. Note that the LIFF module can calculate this pixel-wise prediction for all coordinates in parallel as $\breve{I}^k = \dot{\boldsymbol{F}}^{k} \odot \boldsymbol{f}^k$.

Figure~\ref{fig:LIFF_input} shows a predicted image $\breve{I}^k$ via the LIFF module and its associated residual image $I^k_{res}$ to be compressed for the given reconstructed image $\hat{I}^{k-1}$ of the previous layer $k$-1, and an uncompressed input image $I^k$ (ground truth) in the current layer $k$. Compared to $\hat{I}^{k-1}$ in Figure~\ref{fig:LIFF_input}-(a), $\breve{I}^k$ in Figure~\ref{fig:LIFF_input}-(b) shows much closer result to $I^k$ in Figure~\ref{fig:LIFF_input}-(c), thus leading to a smaller amount of residues $I^k_{res}$ in Figure~\ref{fig:LIFF_input}-(d).

\subsection{Optimization}
\label{subsec:optimization}
We train the whole elements of our COMPASS in an end-to-end manner with the frozen pre-trained image compression module of the BL. To boost up the training, we use the separately pre-trained LIFF and residual compression modules.
To train the COMPASS architecture, we use a combined RD loss function as:
\vspace{-0.1cm}
\begin{equation}
\label{eq:total_loss}
    \begin{aligned}
    L = & \; \sum_{k=1}^K R^k + \lambda \cdot D^k,
    \end{aligned}
\vspace{-0.1cm}
\end{equation}
where $R^k$ and $D^k$ represent a rate term and a distortion term for the EL-$k$, respectively. As in other NN-based image compression methods~\cite{balle2016end, balle2018variational, minnen2018joint, lee2018context}, the rate and distortion are jointly optimized, but we use the summation of those for the $K$ ELs. It should be noted that we use the same $\lambda$ value for the $K$ ELs to maintain the R-D balance over the whole layers. The rate term $R^k$ is the estimated rate amount for the EL-$k$. Specifically, it is determined as the summation of cross-entropy values for latent representations $\boldsymbol{y}^k$ and $\boldsymbol{z}^k$. $\boldsymbol{y}^k$ is the latent representation transformed from an input residual image $I_{res}^k$ via the encoder network of the residual compression module, and $\boldsymbol{z}^k$ is the latent representation transformed from the representation $\boldsymbol{y}^k$ via the hyper-encoder network of the residual compression module, as in the previous hyperprior-based models~\cite{balle2018variational, minnen2018joint, lee2018context}. The rate term $R^k$ is represented as $R^k = H^k(\boldsymbol{\tilde{y}}^k | \boldsymbol{\tilde{z}}^k) + H^k(\boldsymbol{\tilde{z}}^k),$
 where $H^k(\boldsymbol{\tilde{y}}^k | \boldsymbol{\tilde{z}}^k)$ and $H^k(\boldsymbol{\tilde{z}}^k)$ are the cross-entropy terms for noisy latent representations $\boldsymbol{\tilde{y}}^k$ and $\boldsymbol{\tilde{z}}^k$ for the EL-$k$, respectively.
 The cross-entropy values are calculated based on the Gaussian entropy model used in the Mean-scale model~\cite{minnen2018joint}. As in other NN-based image compression methods~\cite{balle2016end, balle2018variational, minnen2018joint, lee2018context}, we sample the noisy latent representations $\boldsymbol{\tilde{y}}^k$ and $\boldsymbol{\tilde{z}}^k$ for the EL-$k$ with the additive uniform noise to fit the samples to the approximate probability mass function (PMF) $P(\cdot)$ of the discretized representations $\boldsymbol{y}^k$ and $\boldsymbol{z}^k$ for the EL-$k$.
$D^k$ is a mean squared error (MSE) between the reconstructed image $\hat{I}^k$ and input image $I^k$ for the EL-$k$.  $\hat{I}^k$ is represented as $\hat{I}^k = \breve{I}^k + \hat{I}_{res}^k$,
where $\hat{I}_{res}^k = D^{RC}(\boldsymbol{\hat{y}}^k)$ and $\boldsymbol{\hat{y}}^{k} = Q(E^{RC}(I_{res}^k))$.
Note that $E^{RC}(\cdot)$ and $D^{RC}(\cdot)$ refer to the encoder and decoder networks of the residual compression module $RC(\cdot)$, respectively, and $Q(\cdot)$ is a rounding function.

Here, we use a rounded latent representation $\boldsymbol{\hat{y}}^k$ rather than the noisy representation $\boldsymbol{\tilde{y}}^k$ for calculating the distortion term. We first used the noisy representation $\boldsymbol{\tilde{y}}^k$ as the input into the decoder network $D^{RC}$, but we obtained very poor optimization results. On the other hand, when we feed the rounded representations $\boldsymbol{\hat{y}}^k$ instead of $\boldsymbol{\tilde{y}}^k$, the coding efficiency is much improved.
The suboptimal performance of the COMPASS with noisy representations could be attributed to the propagation of small errors in the reconstructions, caused by the additive uniform noise, to the following ELs. This propagation of errors could result in a significant discrepancy between the training and inference, ultimately leading to poor results. The hierarchical and recursive operation of the COMPASS may also contribute to this issue.
Whereas, using the rounded representation can certainly prevent the mentioned error propagation because it does not cause any discrepancy between the training and inference phases at all. To deal with the discontinuity due to the rounding operations, we just bypass the gradients backward. In contrast to the distortion term, it should be noted that our COMPASS still uses the noisy representation $\boldsymbol{\tilde{y}}^k$ to calculate the rate term $R^k$ to fit the samples to the approximate PMF $P(\cdot)$. Further details in optimization are described in Appendix A.

\section{Experiments}
\label{sec:Experiments}
We first describe the experimental setup in terms of two aspects: coding efficiency with a fixed scale factor of 2 and coding efficiency with arbitrary scale factors. Then we present the corresponding experimental results.
\begin{table}[!]
{\small
\centering
\begin{tabular}{@{}lclll@{}}
\toprule
\multicolumn{1}{c}{Methods}                   & \multicolumn{3}{c}{BD-rate$\downarrow$} & \multicolumn{1}{c}{Params.}\\ \midrule
SHVC \cite{boyce2015overview}                            & \multicolumn{3}{c}{-33.34\%} & \multicolumn{1}{c}{-} \\
Simulcast (Factorized \cite{balle2016end})            & \multicolumn{3}{c}{-41.71\%} & \multicolumn{1}{c}{-}\\
Simulcast (Mean-scale \cite{minnen2018joint})           & \multicolumn{3}{c}{-22.90\%} & \multicolumn{1}{c}{-}\\
Mei \etal \cite{mei2021learning} (original)         & \multicolumn{3}{c}{-32.12\%} & \multicolumn{1}{c}{40.7M}\\
Mei \etal \cite{mei2021learning} (enhanced)       & \multicolumn{3}{c}{-14.23\%} & \multicolumn{1}{c}{52.8M}\\ 
Single-layer (Mean-scale \cite{minnen2018joint}) & \multicolumn{3}{c}{4.74\%}   & \multicolumn{1}{c}{-}\\ \midrule
COMPASS (proposed) & \multicolumn{3}{c}{-}   & \multicolumn{1}{c}{15.5M}\\ \bottomrule
\end{tabular}
\vspace{-0.2cm}
\caption{Coding efficiency and model size comparison. BD-rate gains of our COMPASS over the various methods are measured in the final EL where the negative values indicate BD-rate gains of our COMPASS. The ‘Params.’ indicates the total number
of parameters.}
\label{tab:coding_efficiency}
}
\end{table}

\begin{figure}[!]
\centering
\includegraphics[scale=0.33]{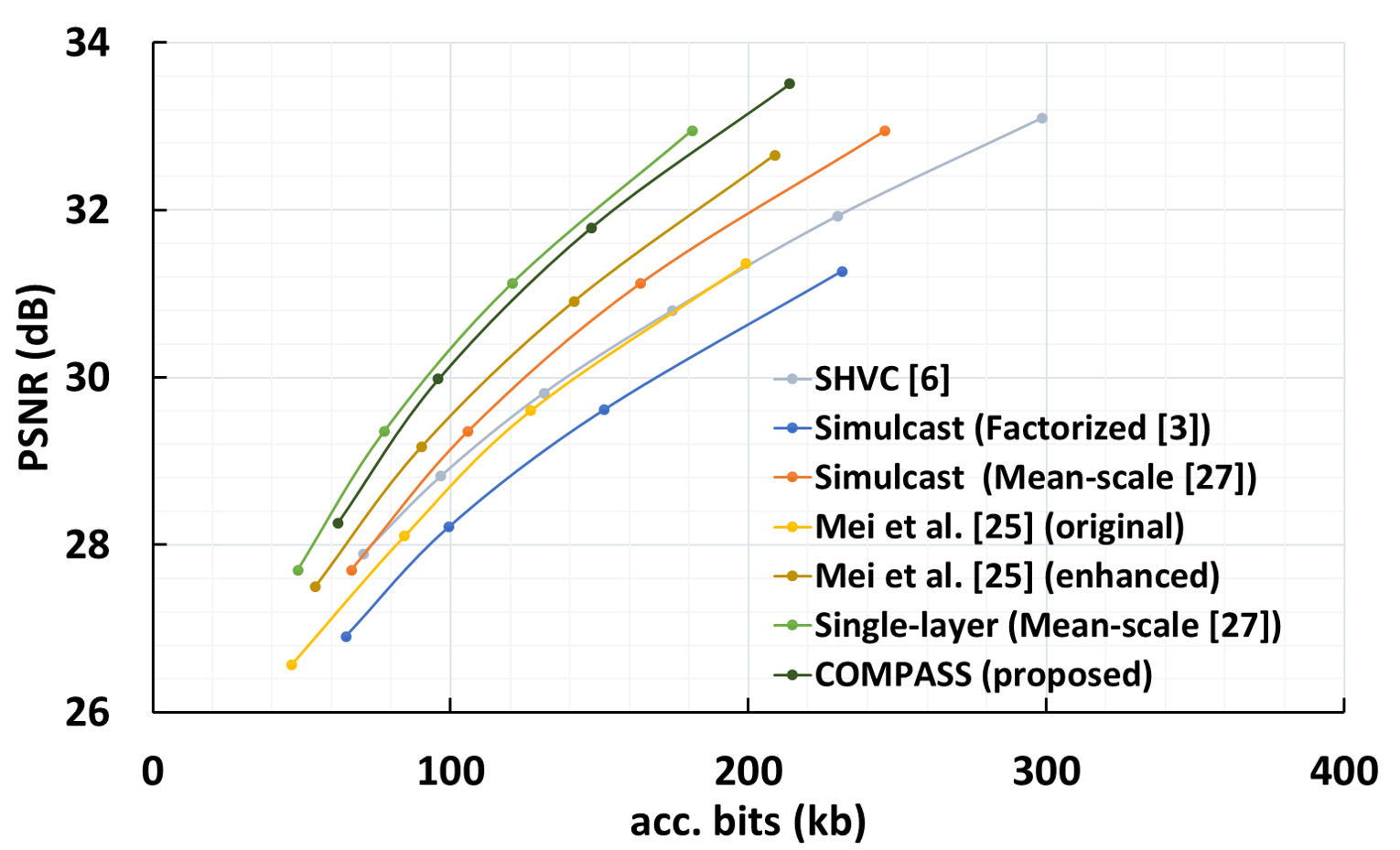}
\caption{The rate-PSNR performance curves of the final ELs for SHVC \cite{boyce2015overview}, the simulcast coding, Mei \etal \cite{mei2021learning}, the single-layer coding, and our COMPASS. The `acc. bits' indicates the accumulated bits up to the final EL.}
\vspace{-0.4cm}
\label{fig:results}
\end{figure}

\begin{table}[!]
{\small
\centering
\renewcommand{\arraystretch}{0.8}
\renewcommand{\tabcolsep}{0.8mm}
\begin{tabular}{@{}lccccc@{}}
\toprule
\multicolumn{1}{c}{\multirow{2}{*}{Methods}} & \multicolumn{5}{c}{Scale Factors (vs. BL)}                    \\ \cmidrule(l){2-6} 
\multicolumn{1}{c}{}                                        & 1.2$\times$      & 1.6$\times$      & 2.0$\times$      & 2.4$\times$      & 2.8$\times$      \\ \midrule
\begin{tabular}[l]{@{}l@{}}SHVC~\cite{boyce2015overview}\end{tabular}                                               & -53.88\% & -42.58\% & -35.87\% & -26.24\% & -22.34\% \\  \midrule
\begin{tabular}[l]{@{}l@{}}Simulcast\\ (Factorized~\cite{balle2016end})\end{tabular}                                 & -53.04\% & -44.87\% & -31.89\% & -34.62\% & -31.59\% \\ \midrule
\begin{tabular}[l]{@{}l@{}}Simulcast\\ (Mean-scale~\cite{minnen2018joint})\end{tabular}                                   & -44.04\% & -31.40\% & -16.29\% & -16.35\% & -12.38\% \\ \midrule
\begin{tabular}[l]{@{}l@{}}Mei \etal \cite{mei2021learning}\\ (original)\end{tabular}                                       & -36.26\% & -29.44\% & -28.20\% & -33.19\% & -36.31\% \\ \midrule
\begin{tabular}[l]{@{}l@{}}Mei \etal \cite{mei2021learning}\\ (enhanced)\end{tabular}                                      & -29.09\% & -17.45\% & -13.52\% & -20.63\% & -23.94\% \\ \midrule
\begin{tabular}[l]{@{}l@{}}Single-layer\\ (Mean-scale~\cite{minnen2018joint})\end{tabular}                                 & -8.19\%  & -3.70\%  & 8.80\%   & 0.31\%   & 0.94\%   \\  \bottomrule
\end{tabular}
\vspace{-0.2cm}
\caption{Coding efficiency comparison for the two-layer scalable coding with arbitrary scale factors. BD-rate gains of our COMPASS over the various methods are measured in the final EL where the negative values indicate BD-rate gains of our COMPASS.}
\vspace{-0.4cm}
\label{tab:configurable_level1}
}
\end{table}

\begin{table}[!]
{\small
\centering
\renewcommand{\arraystretch}{0.8}
\renewcommand{\tabcolsep}{0.8mm}
\begin{tabular}{@{}lccccc@{}}
\toprule
\multicolumn{1}{c}{\multirow{2}{*}{Methods}} & \multicolumn{5}{c}{Scale Factors (vs. BL)}                    \\ \cmidrule(l){2-6} 
\multicolumn{1}{c}{}                                        & 2.4$\times$      & 2.8$\times$      & 3.2$\times$      & 3.6$\times$      & 4.0$\times$      \\ \midrule
SHVC \cite{boyce2015overview}                                                   & -58.33\% & -51.51\% & -46.72\% & -43.65\% & -33.34\% \\ \midrule
\begin{tabular}[l]{@{}l@{}}Simulcast\\ (Factorized~\cite{balle2016end})\end{tabular}                               & -61.04\% & -55.09\% & -50.42\% & -47.11\% & -41.71\% \\ \midrule
\begin{tabular}[l]{@{}l@{}}Simulcast\\ (Mean-scale~\cite{minnen2018joint})\end{tabular}                              & -49.85\% & -42.20\% & -35.33\% & -30.81\% & -22.90\% \\ \midrule
\begin{tabular}[l]{@{}l@{}}Mei \etal \cite{mei2021learning}\\ (original)\end{tabular}                                      & -47.17\% & -38.73\% & -34.34\% & -33.56\% & -32.12\% \\ \midrule 
\begin{tabular}[l]{@{}l@{}}Mei \etal \cite{mei2021learning}\\ (enhanced)\end{tabular}                                      & -38.23\% & -26.46\% & -19.70\% & -17.08\% & -14.23\% \\ \midrule
\begin{tabular}[l]{@{}l@{}}Single-layer\\ (Mean-scale~\cite{minnen2018joint})\end{tabular}                                 & -6.60\%  & -4.23\%  & -1.25\%  & -0.74\%  & 4.74\%   \\ \bottomrule
\end{tabular}
\vspace{-0.2cm}
\caption{Coding efficiency comparison for the three-layer scalable coding with arbitrary scale factors. BD-rate gains of our COMPASS over the various methods are measured in the final EL where the negative values indicate BD-rate gains of our COMPASS. We set the scale factor of the EL-1 relative to the BL to 2, equally.}
\vspace{-0.4cm}
\label{tab:configurable_level2}
}
\end{table}

\subsection{Coding efficiency with a fixed scale factor of 2}
\label{subsec:coding_efficiency}
\noindent \textbf{Experimental setup.}
To validate the coding efficiency of our COMPASS, we compare it with SHVC~\cite{boyce2015overview}, the simulcast coding, Mei~\etal\cite{mei2021learning}, and the single-layer coding with the Mean-scale model~\cite{minnen2018joint}, in terms of BD-rate.
We conduct the coding efficiency comparison of each method for three-layer scalability (one BL and two ELs) with a fixed scale factor of 2 between adjacent layers.
For a fair comparison, we used both Mei~\etal\cite{mei2021learning}'s original version with the Factorized model~\cite{balle2016end} and its enhanced one with the Mean-scale model~\cite{minnen2018joint} as the same image compression model as ours. Note that we set the same numbers of channels for all image compression modules with $N$ = $128$ and $M$ = $192$ \footnote{$N$ and $M$ refer to the output channels of the encoder network and hyper-encoder network of the image compression modules, respectively.}. 
Following the typical validation procedures \cite{boyce2015overview, seregin2014common} for the scalable coding, we measure the BD-rate performance for the final ELs with the largest scales (spatial sizes) of reconstructions, but we also provide the BD-rate results for the BL and intermediate ELs in Appendix B.
For the simulcast and single-layer coding, we use the pre-trained models from CompressAI-Pytorch~\cite{begaint2020compressai}.
The BD-rate performance of SHVC~\cite{boyce2015overview} is measured using the All-Intra Mode of the SHVC reference software (SHM-12.4)~\cite{SHM} with the QP values of 30, 32, 34, 36, 38 and 40. The RGB inputs are converted into the YUV420 format and the reconstructions are converted back to the RGB format to achieve the best possible performance of SHVC~\cite{boyce2015overview}.
More specifically, we use the Kodak Lossless True Color Image dataset \cite{franzen1999kodak} that consists of 24 768$\times$512-sized (or 512$\times$768-sized) images. For the downscaling of the input images, we use the bicubic interpolation function implemented in Pytorch \cite{paszke2019pytorch}. For the feature extractor of the LIFF module in our COMPASS, we set the number of residual dense blocks (RDBs) and convolutional layers of each RDB to 4. The number of output channels and the growth rate of each RDB are set to 64 and 32, respectively. For the filter generation MLP of the LIFF module, we set the number of hidden layers to 5, each of which has 256 output channels. The COMPASS is built using the open-source CompressAI Pytorch library \cite{begaint2020compressai}.

\noindent \textbf{Experimental results.}
Table \ref{tab:coding_efficiency} shows the coding efficiency performance in terms of BD-rate for our COMPASS against the compared methods. It should be noted in Table \ref{tab:coding_efficiency} that each compared method becomes an anchor for comparison against our COMPASS. Therefore, the negative percentage values indicate that our COMPASS outperforms the corresponding methods in BD-rate coding efficiency by those amounts while the positive values imply under-performance of the COMPASS. Figure~\ref{fig:results} shows the rate-PSNR performance curves for Table \ref{tab:coding_efficiency}. As shown in Table \ref{tab:coding_efficiency} and Figure~\ref{fig:results}, our COMPASS significantly outperforms all the spatially scalable coding methods except the Single-layer (Mean-scale~\cite{minnen2018joint}). It is worthy to note that, although that Mei \etal \cite{mei2021learning}'s enhanced version uses the same image compression module (Mean-scale~\cite{minnen2018joint}) as the COMPASS and focuses only on the fixed scale factor of 2, -14.23\% of BD-rate gain is achieved by our COMPASS that supports various different scale factors, which will be discussed in Sec. \ref{subsec:scale_configurability}.
It is noted in Table \ref{tab:coding_efficiency} that our COMPASS has a less number of parameters, compared to Mei \etal \cite{mei2021learning}'s method.
Furthermore, impressively, our COMPASS achieves comparable results to the single-layer coding with the Mean-scale model~\cite{minnen2018joint}, while the existing scalable coding methods \cite{boyce2015overview} show considerably lower coding efficiency compared with their single-layer coding as described in \cite{boyce2015overview, seregin2014common} due to low inter-layer prediction accuracy. In addition, our COMPASS shows the BD-rate gains of -24.97\% and -35.87\% compared to SHVC~\cite{boyce2015overview} at the BL and the EL-1, respectively. Further comparisons for the other layers are provided in Appendix B.

\subsection{Coding efficiency with arbitrary scale factors}
\label{subsec:scale_configurability}
\noindent \textbf{Experimental setup.}
We show the effective coding efficiency of our COMPASS at arbitrary scales by comparing it with the six  coding methods. For this, five scale factors are used for each of two experiments. The first experiment is conducted with five two-layer scalabilities (one BL and one EL (EL-1: 1.2$\times$, 1.6$\times$, 2.0$\times$, 2.4$\times$, and 2.8$\times$)) while the second one is with five three-layer scalabilities (one BL and two ELs (EL-1: 2.0$\times$ and EL-2: 2.4$\times$, 2.8$\times$, 3.2$\times$, 3.6$\times$, and 4.0$\times$)). More experiment results are provided for more combinations of scale factors in Appendix B.
For Mei~\etal\cite{mei2021learning} which only supports a fixed scale factor of 2 between adjacent layers, we upscale or downscale the output images using bicubic interpolation to match with other scale factors. The other experimental conditions such as datasets and channel numbers are the same as those in Sec. \ref{subsec:coding_efficiency}.

\noindent \textbf{Experimental results.}
Table \ref{tab:configurable_level1} shows the comparison results of the two-layer scalable coding. As shown, our COMPASS significantly outperforms all the spatially scalable coding methods over the entire range of the different scale factors from 1.2$\times$ to 2.8$\times$. Furthermore, it achieves comparable results to the single-layer coding with the Mean-scale model~\cite{minnen2018joint} over the entire range. Surprisingly, our COMPASS even outperforms it for the scale factors of 1.2$\times$ and 1.6$\times$.
This is because the input images for the single-layer coding need to be padded to a multiple of 64 in order to be processed into the CNN architecture of the image compression network, which can lead to lower the coding efficiency. In contrast, our LIFF module is effective at handling the arbitrary scale factors.
For the three-layer scalable coding, as shown in Table \ref{tab:configurable_level2}, our COMPASS also outperforms all the spatially scalable coding methods as the two-layer scalable coding, and even exhibits a superiority to the single-layer coding with the Mean-scale model~\cite{minnen2018joint} over the whole scale factors only except scale factor 4.0$\times$.
The superiority of our COMPASS stems from the fact that the LIFF module can well perform the inter-layer prediction for arbitrary scale factors.

\begin{figure*}[h]
\centering
\includegraphics[scale=0.45]{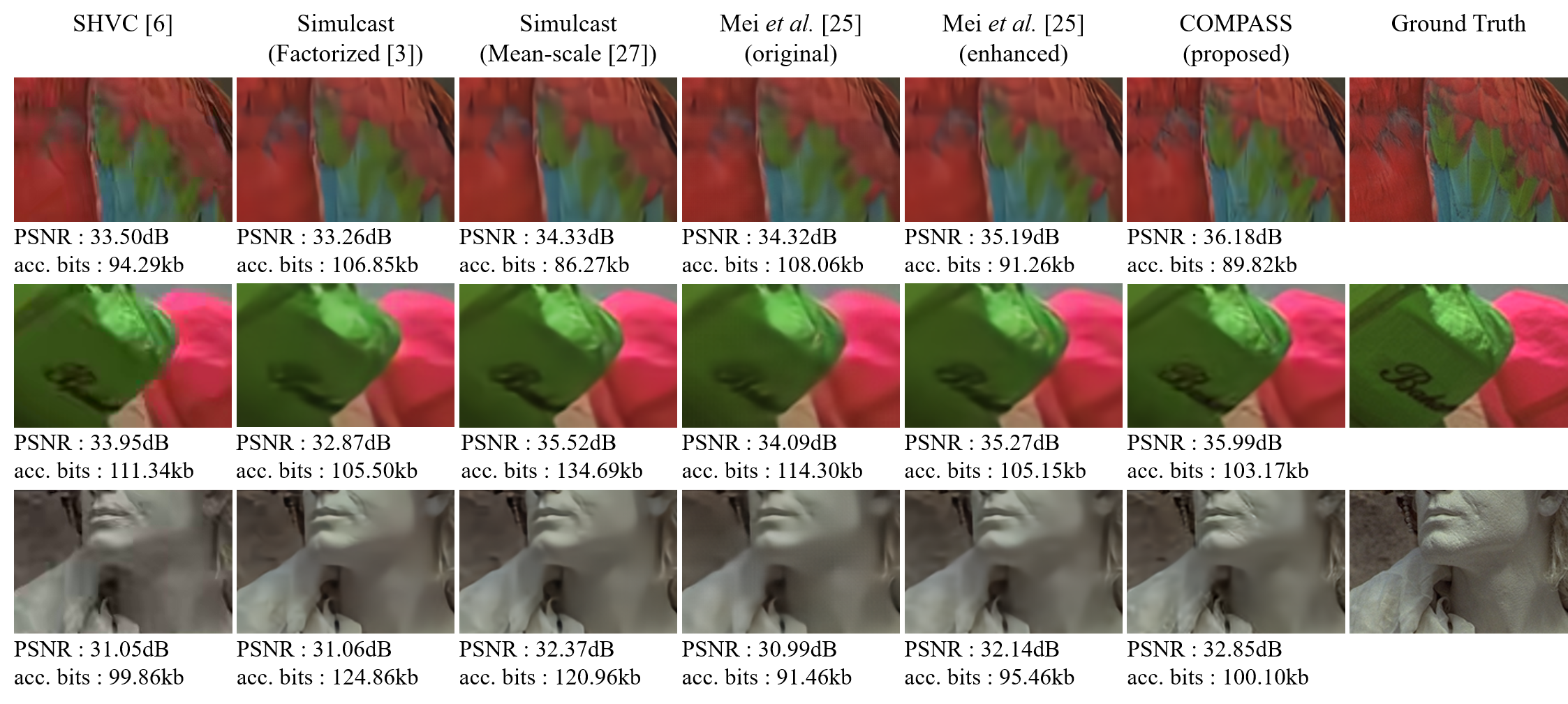}
\vspace{-0.2cm}
\caption{Visual comparison results for \textit{kodim23.png}, \textit{kodim03.png}, \textit{kodim17.png} images in Kodak Lossless True Color
Image dataset~\cite{franzen1999kodak} (best viewed in digital format). The `acc. bits' indicates the accumulated bits up to the final EL. We match the accumulated bits among the compared methods as much as possible. \textbf{Zoom for better visual comparison}.}
\vspace{-0.4cm}
\label{fig:visual_comparison}
\end{figure*}

\subsection{Visual comparison}
Figure \ref{fig:visual_comparison} shows the visual comparison results of our COMPASS with SHVC~\cite{boyce2015overview}, the simulcast coding, and Mei \etal \cite{mei2021learning} for the three-layer scalable coding with a scale factor of 2 between adjacent layers. The images shown in Figure \ref{fig:visual_comparison} are the largest reconstructions obtained from the final EL. We set the accumulated bits as close as possible between the methods. As shown, the subjective qualities of the reconstructions by our COMPASS are significantly better than those from the other methods, and especially, the high-frequency components such as edges and textures are more clearly reconstructed. More visual comparison results are provided in Appendix C, where we also provide the reconstructions with the multi-layer configuration greater than three layers to verify the extensibility of our COMPASS in terms of the number of layers.

\subsection{Ablation study}
To verify the effectiveness of the proposed elements (or optimization strategy) in our COMPASS, we measure the coding efficiency of the ablated models and compare the results with those of the full model of our COMPASS in terms of BD-rate. In the comparison, the ablated elements are the LIFF module described in Sec. \ref{subsec:LIFF}, the convolutional-layer-wise padding described in Sec. \ref{subsec:overall_architecture}, and the adoption of rounded representations replacing the noisy representations for calculation of the distortion term in training described in Sec. \ref{subsec:optimization}. It should be noted for the ablated model without the LIFF module that a bicubic interpolation is used instead. For the ablated model without the convolutional-layer-wise padding, we pad the input images to match their sizes to the multiple of $64$ in both vertical and horizontal axes. For the ablated model without using the rounded representations in training, we use the noisy representations instead of the rounded representations. As shown in Table \ref{tab:ablation_study}, our full COMPASS model significantly outperforms all the ablated models, which represents that each proposed element of the COMPASS model effectively contributes to the enhancement of coding efficiency.


\begin{table}[!]
{\small
\centering
\renewcommand{\arraystretch}{0.8}
\begin{tabular}{@{}ccc|c@{}}
\toprule
\begin{tabular}[c]{@{}c@{}}Inter-layer\\ prediction\end{tabular} & \begin{tabular}[c]{@{}c@{}}Conv. layer \\ padding\end{tabular}& \begin{tabular}[c]{@{}c@{}}Rounded\\ latent rep.\end{tabular} & \begin{tabular}[c]{@{}c@{}}BD-rate$\downarrow$\\ (vs. full model)\end{tabular} \\ \midrule 
 Bicubic&\cmark&\cmark& 9.27\% \\ 
 LIFF&\xmark&\cmark& 18.95\% \\ 
 LIFF&\cmark&\xmark& 13.00\% \\ \bottomrule
\end{tabular}
\vspace{-0.2cm}
\caption{Ablation study results about the proposed elements or optimization strategies in our COMPASS.}
\vspace{-0.5cm}
\label{tab:ablation_study}
}
\end{table}

\section{Conclusion}
In this paper, we propose a new NN-based spatially scalable image compression method, called COMPASS, which can achieve high coding efficiency while supporting arbitrary scale factors between adjacent layers, not limited to doubling the scales. To be an effective architecture, all the enhancement layers share the LIFF module and the residual compression module, which are recursively performed into higher scale factors. The LIFF module is adopted for the inter-layer arbitrary scale prediction, which can effectively reduce the spatial redundancy between layers for arbitrary scale factors. We also propose the combined RD loss function to effectively train multiple layers. Experimental results show that our COMPASS significantly outperforms SHVC~\cite{boyce2015overview}, the simulcast coding, and the existing NN-based spatially scalable coding method~\cite{mei2021learning} in terms of BD-rate for all combinations of scale factors. Our COMPASS also uses a smaller number of parameters than the existing NN-based spatially scalable coding method~\cite{mei2021learning}. To the best of our knowledge, the COMPASS is the first work that shows comparable or even better performance in terms of coding efficiency than the single-layer coding for various scale factors, based on a same image compression backbone.

\section*{Acknowledgement}
This work was supported by internal fund/grant of Electronics and Telecommunications Research Institute (ETRI). [23YC1100, Technology Development for Strengthening Competitiveness in Standard IPR for communication and media]

{\small
\bibliographystyle{ieee_fullname}
\bibliography{main}
}

\end{document}